\documentclass[cits]{PoS}

\providecommand{\href}[2]{#2}
\addcontentsline{toc}{section}{References}
\usepackage{pifont}

\def\Poles{{\cal P}oles}
\def\dsigma{{\rm d} \hat\sigma}

\title{Infrared structure of pp $\to$ 2 jets at NNLO: the gluon channel}

\ShortTitle{Infrared structure of pp $\to$ 2 jets at NNLO: the gluon channel}

\author{\speaker{Joao Pires}%
        \\
        Institute for Theoretical Physics, ETH Zurich, 8093 Zurich, Switzerland\\
       E-mail: \email{pires@itp.phys.ethz.ch}}

\abstract{We use the antenna subtraction method to isolate the infrared singularities present in QCD scattering amplitudes at next-to-next-to-leading order. In particular, infrared singularities
due to double-real radiation and real-virtual radiation are subtracted from the QCD matrix elements using antenna functions which are then integrated analytically and added to the
double-virtual contribution. Here we consider two-jet production at NNLO at hadron colliders and construct subtraction terms for the double-real and real-virtual channels
that describe the single and double unresolved configurations of the pure gluon scattering matrix elements. In all singular regions we show numerically that the subtraction terms correctly
approximate the matrix elements and demonstrate that upon integration they contribute 
to the cancellation of all infrared poles when combined with one and two-loop matrix elements.}

\FullConference{``Loops and Legs in Quantum Field Theory''  11th DESY Workshop on Elementary Particle Physics \\
		 April 15-20, 2012\\
		 Wernigerode, Germany}

\begin{document}

\section{Introduction}
In high energy hadron-hadron collisions, jet production is the dominant high transverse momentum process.
It results from a scattering of a coloured parton off another coloured parton, a process whose cross section is described in the following factorised form, 

\begin{equation}
\label{eq:totsig}
{\rm d}\sigma =\sum_{i,j} \int   
\frac{d\xi_1}{\xi_1} \frac{d\xi_2}{\xi_2} f_i(\xi_1,\mu_F) f_j(\xi_2,\mu_F) \dsigma_{ij}(\alpha_s(\mu_R),\mu_R,\mu_F). \nonumber
\end{equation}
In (\ref{eq:totsig}) the probability of finding a parton of type $i$ in the proton carrying a momentum fraction $\xi$ is described by the parton distribution function $f_i(\xi,\mu_F)d\xi$ and the parton-level scattering cross section ${\rm d}\hat\sigma_{ij}$  for parton $i$ to scatter off parton $j$ normalised to the hadron-hadron flux\footnote{The partonic cross section normalised to the parton-parton flux is obtained by absorbing the inverse factors of $\xi_1$ and $\xi_2$ into $\dsigma_{ij}$.} is summed over the possible parton types $i$ and $j$. As usual $\mu_R$ and $\mu_F$ are the renormalisation and factorisation scales. 

The inclusive jet $pp\to j +X$ and dijet $pp\to jj+X$ cross sections have been studied at the TEVATRON \cite{tevatron} and at the LHC \cite{LHC} where the advantage of very high statistics and steady improved control over systematic uncertainties \cite{Aad:2011he} allows these measurements to be performed with a few percent accuracy. This high quality jet data which probes the dynamic of the hard scattering simultaneously at a wide range of momentum transfers thus contains precision information on the structure of the proton and on the QCD coupling constant $\alpha_s$. However, extraction of these fundamental parameters from the data depends on reliable theoretical calculations which should match the precision of the experimental measurement.

The uncertainties related to the theoretical prediction may be improved by including 
higher order perturbative corrections which have the effect of (a) reducing the renormalisation/factorisation scale dependence of the cross section and (b) improving the matching of the parton level event topology with the experimentally observed hadronic final state~\cite{Glover:2002gz}.
In this way the partonic cross section ${\rm d}\hat\sigma_{ij}$ in~(\ref{eq:totsig}) has the perturbative expansion 
\begin{equation}
{\rm d}\hat\sigma_{ij}={\rm d}\hat\sigma_{ij}^{LO}
+\left(\frac{\alpha_s}{2\pi}\right){\rm d}\hat\sigma_{ij}^{NLO}
+\left(\frac{\alpha_s}{2\pi}\right)^2{\rm d}\hat\sigma_{ij}^{NNLO}
+{\cal O}(\alpha_s^3)
\end{equation}
where the next-to-leading order (NLO) and next-to-next-to-leading order (NNLO) strong corrections are identified. The NLO corrections for single inclusive jet and dijet distributions have been computed by several groups~\cite{NLOinclusivejets} and have been successfully compared with data from the TEVATRON~\cite{tevatron} and LHC~\cite{LHC}. 
In this talk we are mainly concerned with the pure gluon channel contribution to the NNLO corrections to dijet production where the resulting theoretical uncertainty is estimated to be at the few per-cent level \cite{Glover:2002gz}.

\section{NNLO calculations}
At NNLO, there are three distinct contributions due to double-real radiation ${\rm{d}}\sigma_{NNLO}^{RR}$, mixed real-virtual radiation ${\rm{d}}\sigma_{NNLO}^{RV}$ and double-virtual radiation ${\rm{d}}\sigma_{NNLO}^{VV}$, that are given by
\begin{eqnarray}
{\rm d}\hat\sigma_{NNLO}&=&\int_{{\rm{d}}\Phi_{m+2}} {\rm{d}}\hat\sigma_{NNLO}^{RR} 
+\int_{{\rm{d}}\Phi_{m+1}} {\rm{d}}\hat\sigma_{NNLO}^{RV} 
+\int_{{\rm{d}}\Phi_m}{\rm{d}}\hat\sigma_{NNLO}^{VV}
\label{eq:NNLOcorr}
\end{eqnarray}
where the integration is over the appropriate $N$-particle final state subject to the constraint that precisely $m$-jets are observed,
\begin{equation}
\int_{{\rm{d}}\Phi_{N}} = \int {\rm{d}}\Phi_{N} J_m^{(N)}.
\end{equation}
For dijet production in the pure gluon channel these contributions read,
\begin{eqnarray}
{\rm d}\hat\sigma_{NNLO}^{RR}&=&N\int{\rm d}\Phi_{4}(p_{3},p_{4},p_{5},p_{6};p_{1},p_{2})
|{\cal M}_{gg\to gggg}^{(0)}|^2\,J_{2}^{(4)}(p_{3},p_{4},p_{5},p_{6})\nonumber\\
{\rm d}\hat\sigma_{NNLO}^{RV}&=&N\int{\rm d}\Phi_{3}(p_{3},p_{4},p_{5};p_{1},p_{2})
\Big({\cal M}_{gg\to ggg}^{(0)*}{\cal M}_{gg\to ggg}^{(1)}+
{\cal M}_{gg\to ggg}^{(0)}{\cal M}_{gg\to ggg}^{(1)*}\Big)\,J_{2}^{(3)}(p_{3},p_{4},p_{5})\nonumber\\
{\rm d}\hat\sigma_{NNLO}^{VV}&=&N\int{\rm d}\Phi_{2}(p_{3},p_{4};p_{1},p_{2})
\Big({\cal M}_{gg\to gg}^{(2)*}{\cal M}_{gg\to gg}^{(0)}+
{\cal M}_{gg\to gg}^{(0)}{\cal M}_{gg\to gg}^{(2)*}+|{\cal M}_{gg\to gg}^{(1)}|^2\Big)\,J_{2}^{(2)}(p_{3},p_{4}),\nonumber
\end{eqnarray}
where ${\cal M}^{(i)}$ denotes the $i$-loop amplitude and the jet function $J_m^{(n)}$ defines the procedure for building $m$-jets from $n$-final state partons.

Individually the parton channels defined above are all separately infrared divergent although, after renormalisation and factorisation of initial state singularities, their sum is finite~\cite{KLN,Ellis:1978ty}. 
The singularities in the double-real channel arise in the single and double unresolved regions of the phase space which physically correspond to two gluons becoming simultaneously soft and/or collinear. In these regions
of the phase space the tree-level matrix element diverges in universal and process independent ways~\cite{doubleunresolved}. 
The real-virtual contribution contains, on the one hand, explicit infrared divergences coming from integrating over the loop momentum and, on the other
hand, implicit poles in the regions of the phase space where one of the final state partons becomes unresolved.  This corresponds to the soft and collinear limits of the one-loop amplitude which again have universal
and process independent limits~\cite{singleunresolved}. 
Finally the double-virtual contribution contains the renormalised two-loop corrections containing explicit infrared poles coming from integrating over the loop momenta~\cite{Catani:1998bh}.

We note that explicit expressions for the interference of the four-gluon tree-level and two-loop amplitudes are available in Refs.~\cite{twoloop4g}, while the self interference of the four-gluon one-loop amplitude
is given in \cite{Glover:2001rd}. The one-loop helicity amplitudes for the five gluon amplitude are given in \cite{Bern:1993mq} and the double-real six-gluon matrix elements were derived in \cite{6g0}. 
All matrix elements relevant for the computation of $gg \to gg$ at NNLO are therefore known. 

The bottleneck for the computation of jet observables at NNLO in this case is that the real and virtual contributions have a different number of final state partons and
require the phase space integrations to be done separately. As a result one is left with the task of finding a systematic procedure to extract the infrared soft and collinear singularities that
are present in the double-real radiation and mixed real-virtual contributions at NNLO before a finite physical prediction can be obtained.

Subtraction schemes are a well established solution to this problem and several methods for systematically constructing general subtraction 
terms have been proposed in the literature at NLO~\cite{NLOschemes} and, with various degrees of 
sophistication, at NNLO~\cite{NNLOschemes,GehrmannDeRidder:2005cm,Glover:2010im,GehrmannDeRidder:2011aa}. 
Another NNLO subtraction scheme has been proposed in~\cite{Catani:2007vq}. 
It is not a general subtraction scheme, but can nevertheless deal with an entire class of processes, those without coloured final states, in hadron-hadron collisions.

In addition, there are purely numerical methods~\cite{SDvirtual,SDreal,Anastasiou:2010pw} such as the completely independent approach called sector decomposition,
which avoids the need for analytical integrations and which has been developed for virtual \cite{SDvirtual} 
and real radiation \cite{SDreal} corrections to NNLO. Finally, a new numerical method combining the ideas of subtraction and sector decomposition has been proposed and developed in~\cite{NNLOFKS}. 
We will follow the NNLO antenna subtraction method which was proposed in \cite{GehrmannDeRidder:2005cm}.

\section{Antenna subtraction}
The subtraction approach is based on the universal factorisation of the QCD matrix elements (as described in the previous section) and of the phase space 
in the singular regions. 

The factorisation property of the matrix elements allows us to find simple counterterms for the real-radiation channels which reproduce the limits of the amplitudes in the different singular kinematic configurations and therefore
regulate the numerical integration of the real contribution. The phase space factorisation property on the other hand allows us to carry out the analytic integration of the countertem such that it can be added to the
virtual contribution in integrated form leading to an analytic pole cancellation between the two. 
This analytic integration is fully inclusive over the unresolved configurations and can be carried out as it is not subject to any jet defining cuts, and does not depend
on the details of the process under consideration. The antenna subtraction scheme is then by construction a general subtraction scheme and, as we will describe below, can be applied to any NNLO QCD 
calculations for exclusive jet observables at hadron colliders.

Suppressing the labels of the partons in the initial state of the hard scattering and introducing subtraction terms, equation (\ref{eq:NNLOcorr}) can be rewritten as~\cite{GehrmannDeRidder:2005cm}:
\begin{eqnarray}
\dsigma_{NNLO}&=&\int_{{\rm{d}}\Phi_{m+2}}\left(\dsigma_{NNLO}^{RR}-\dsigma_{NNLO}^S\right)
+\int_{{\rm{d}}\Phi_{m+2}}\dsigma_{NNLO}^S\nonumber\\
&+&\int_{{\rm{d}}\Phi_{m+1}}\left(\dsigma_{NNLO}^{RV}-\dsigma_{NNLO}^{VS}\right)
+\int_{{\rm{d}}\Phi_{m+1}}\dsigma_{NNLO}^{VS}
+\int_{{\rm{d}}\Phi_{m+1}}\dsigma_{NNLO}^{MF,1}\nonumber\\
&+&\int_{{\rm{d}}\Phi_m}\dsigma_{NNLO}^{VV}
+\int_{{\rm{d}}\Phi_m}\dsigma_{NNLO}^{MF,2}.
\end{eqnarray}
Here, $\dsigma^{S}_{NNLO}$ denotes the subtraction term for the $(m+2)$-parton final state which behaves like the double-real radiation contribution
$\dsigma^{RR}_{NNLO}$ in all singular limits. 
Likewise, $\dsigma^{VS}_{NNLO}$ is the one-loop virtual subtraction term 
coinciding with the one-loop $(m+1)$-final state $\dsigma^{RV}_{NNLO}$ in all singular limits. 
The two-loop correction 
to the $m$-parton final state is denoted by $\dsigma^{VV}_{NNLO}$.  In addition, as there are partons in the initial state, 
there are also two mass factorisation contributions, 
$\dsigma^{MF,1}_{NNLO}$ and $\dsigma^{MF,2}_{NNLO}$, for the $(m+1)$- and $m$-particle final states respectively.

In order to construct a numerical implementation of the NNLO cross section, 
the various contributions must be reorganised according to the number of final state particles,
\begin{eqnarray}
\dsigma_{NNLO}&=&\int_{{\rm{d}}\Phi_{m+2}}\left[\dsigma_{NNLO}^{RR}-\dsigma_{NNLO}^S\right]
\nonumber \\
&+& \int_{{\rm{d}}\Phi_{m+1}}
\left[
\dsigma_{NNLO}^{RV}-\dsigma_{NNLO}^{T}
\right] \nonumber \\
&+&\int_{{\rm{d}}\Phi_{m\phantom{+1}}}\left[
\dsigma_{NNLO}^{VV}-\dsigma_{NNLO}^{U}\right],
\label{eq:antSub}
\end{eqnarray}
where the terms in each of the square brackets are finite and well behaved in the infrared singular regions.  More precisely,
\begin{eqnarray}
\label{eq:Tdef}
\dsigma_{NNLO}^{T} &=& \phantom{ -\int_1 }\dsigma_{NNLO}^{VS}
- \int_1 \dsigma_{NNLO}^{S,1} - \dsigma_{NNLO}^{MF,1},  \\
\label{eq:Udef}
\dsigma_{NNLO}^{U} &=& -\int_1 \dsigma_{NNLO}^{VS}
-\int_2 \dsigma_{NNLO}^{S,2}
-\dsigma_{NNLO}^{MF,2}.
\end{eqnarray}
where the double-real integrated subtraction term gives contributions to both $(m+1)$- and $m$-parton final states,
\begin{equation}
\int_{{\rm{d}}\Phi_{m+2}}\dsigma_{NNLO}^S
= \int_{{\rm{d}}\Phi_{m+1}} \int_1 \dsigma_{NNLO}^{S,1}
+\int_{{\rm{d}}\Phi_{m}} \int_2 \dsigma_{NNLO}^{S,2}.
\end{equation} 

The derivation and specific forms of the $\dsigma^{S}_{NNLO}$ and $\dsigma^{VS}_{NNLO}$ counterterms for the case of coloured particles in the final state has been discussed in~\cite{GehrmannDeRidder:2005cm}. 
The extension to the case of coloured particles in the initial state relevant for studying processes at hadron colliders is described in~\cite{Glover:2010im,GehrmannDeRidder:2011aa}.

In a brief description,
the subtraction terms are constructed from products of antenna functions with reduced matrix elements (with fewer final state partons than the original matrix element). The relevant antennae are
determined by both the external state and the pair of hard
partons it collapses to. In general we denote the antenna function as $X$. For antennae
that collapse onto a hard quark-antiquark pair, $X = A$ for $qg\bar{q}$. Similarly, for quark-gluon antenna, we have $X = D$ for $qgg$ and $X = E$ for $qq\bar{q}$ final states.
Finally, we characterise the gluon-gluon antennae as $X = F$ for $ggg$, $X=G$ for $gq\bar{q}$ final states. In all cases the antenna functions are derived from physical matrix elements: 
the quark-antiquark antenna functions from 
$\gamma^* \to q\bar q~+$~(partons)~\cite{GehrmannDeRidder:2004tv}, the quark-gluon antenna 
functions from $\tilde\chi \to \tilde g~+$~(partons)~\cite{GehrmannDeRidder:2005hi} and 
the gluon-gluon antenna functions from $H\to$~(partons)~\cite{GehrmannDeRidder:2005aw}. The full subtraction term is obtained by summing over all antennae required for the problem under consideration. In the most general
case (two partons in the initial state, and two or more hard partons in the final state), this sum includes final-final, initial-final and initial-initial antennae. The antenna subtraction formalism has also been
extended to deal with hadronic observables involving massive fermions at NNLO~\cite{massiveantennae}.

A key element in any subtraction method, is the ability to add back the subtraction term integrated over the unresolved phase space. The subtraction terms are then integrated over a phase space which
is factorised into an antenna phase space (involving all unresolved partons and the two radiators) multiplied with a reduced phase space (where the momenta of radiators and unesolved radiation
are replaced by two redefined momenta).  In the antenna approach,
this integration needs to be performed once for each antenna and the analytic integration of those is 
performed over the antenna phase space only. All the necessary integrals relevant for jet cross sections at NNLO at hadron-hadron colliders 
have been computed and are summarised in Table 1.

\begin{table*}[tbh]
\begin{center}
\caption{Integrated massless antenna functions}
\label{table:1}
\renewcommand{\tabcolsep}{2pc} 
\renewcommand{\arraystretch}{1.2} 
\begin{tabular}{|c|c|c|c|}
\hline
         & {\rm final-final} & {\rm final-initial} & {\rm initial-initial}   \\
\hline
$X_3^0$                &  \ding{51}\cite{GehrmannDeRidder:2005cm} & \ding{51}\cite{Daleo:2006xa}  &  \ding{51}\cite{Daleo:2006xa}     \\
$X_4^0$                &  \ding{51}\cite{GehrmannDeRidder:2005cm} & \ding{51}\cite{Daleo:2009yj}  &   \ding{51} \cite{iiNNLOantennae}  \\
$X_3^0\otimes X_3^0$   &  \ding{51}\cite{GehrmannDeRidder:2005cm} &  \ding{51}\cite{VV} &   \ding{51}\cite{VV} \\
$X_3^1$  	       & \ding{51}\cite{GehrmannDeRidder:2005cm} &  \ding{51}\cite{Daleo:2009yj} &  \ding{51}\cite{Gehrmann:2011wi}   \\
\hline
\end{tabular}\\[2pt]
\end{center}
\end{table*}

\section{Dijet production at NNLO at hadron colliders}
In this section we will discuss the ongoing calculation of dijet production at NNLO at hadron colliders using the antenna subtraction scheme.
\subsection{Double-real contribution}
\begin{figure}[ht]
\begin{minipage}[b]{0.3\linewidth}
\centering
\includegraphics[width=4cm]{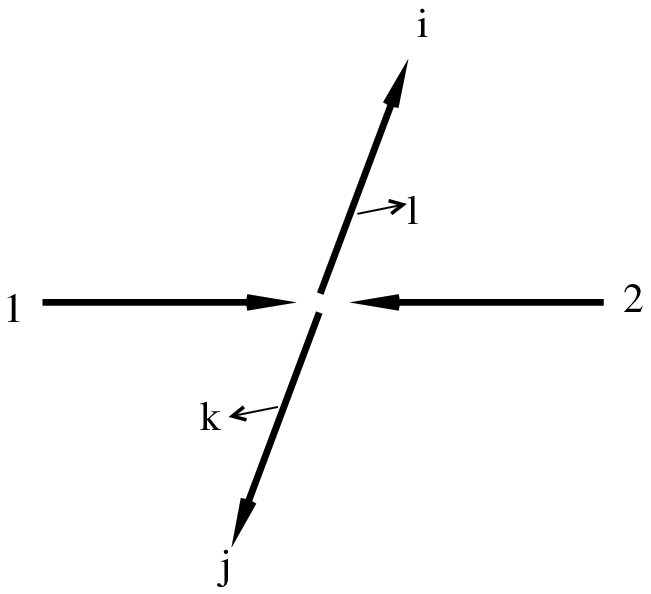}\\
\vspace{0.2cm}
(a)
\end{minipage}
\hspace{0.5cm}
\begin{minipage}[b]{0.7\linewidth}
\centering
\includegraphics[width=5cm,angle=270]{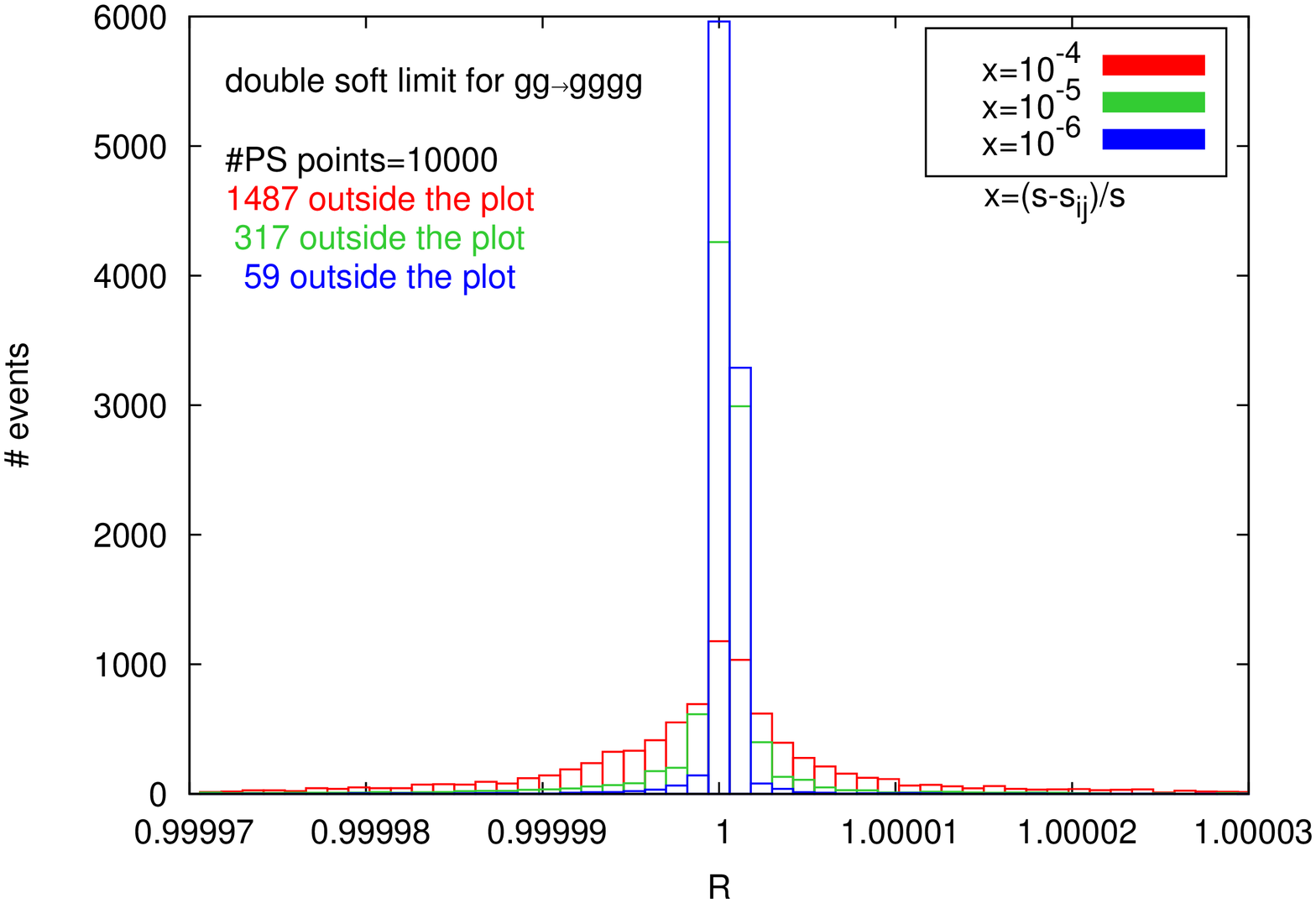}\\
\vspace{0.2cm}
(b)
\end{minipage}
\caption[Double soft limit distributions]{(a) Example configuration of a double soft event in the double-real channel with $s_{ij}\approx s_{12}=s$.
(b) Distribution of $R_{RR}$ for 10000 double soft phase space points.}
\label{fig:dsoft}
\end{figure}

In a previous paper~\cite{Glover:2010im}, the subtraction term $\dsigma_{NNLO}^S$ in (\ref{eq:antSub}), corresponding to the leading colour pure gluon contribution to dijet production at hadron colliders was derived. 
The subtraction term was shown to reproduce the singular behaviour present in $\dsigma_{NNLO}^{RR}$ in all of the single and double unresolved limits. One example of this can be seen in fig.~\ref{fig:dsoft}(b) where
we generated 10000 random double soft phase space points and show the distribution of the ratio between the matrix element and the subtraction term $R_{RR}$. The three colours
represent different values of $x=(s-s_{ij})/s$ [$x=10^{-4}$ (red), $x=10^{-5}$ (green), $x=10^{-6}$ (blue)] and we can see that for smaller values of $x$ we go closer to the singular region
and the distribution peaks more sharply around unity.

We also analysed the regions of phase space corresponding to the final and initial state single collinear singularities. There we saw that the subtraction term,
which is based on azimuthally averaged antenna functions does not accurately describe the azimuthal correlations present in the matrix elements
and antenna functions when a gluon splits into two collinear gluons. Nevertheless, the azimuthal terms coming from the single collinear limits do vanish after an azimuthal integration 
over the unresolved phase space with their angular dependence being of the form $A\cos(2\phi+\alpha)$, where $\phi$ is the azimuthal angle of the collinear system.
Therefore by combining two phase space points with azimuthal angles $\phi$ and $\phi+\frac{\pi}{2}$~\cite{Glover:2010im,Pires:2010jv} and all other coordinates equal, the azimuthal correlations drop out since
the correlations present in the matrix element at the rotated point cancel precisely the azimuthal correlations of the un-rotated point.

\subsection{Real-virtual contribution}
In a subsequent paper~\cite{GehrmannDeRidder:2011aa} we derived the subtraction term $\dsigma_{NNLO}^T$ in (\ref{eq:antSub}) corresponding
to the leading colour pure gluon contribution to dijet production at hadron colliders. In~\cite{GehrmannDeRidder:2011aa} we established which contributions from the double-real subtraction term have to be
added back in integrated form at the $(m+1)$ and $m$-parton level. By construction the counterterm removes the explicit infrared poles present
on the one-loop amplitude, as well as the implicit singularities that occur in the soft and collinear limits. The $\epsilon$-poles present in the real-virtual contribution
are analytically cancelled by the $\epsilon$-poles in the subtraction term rendering the real-virtual contribution locally finite over the whole of phase space.

\begin{figure}[t]
\begin{minipage}[b]{0.3\linewidth}
\centering
\includegraphics[width=5cm,angle=270]{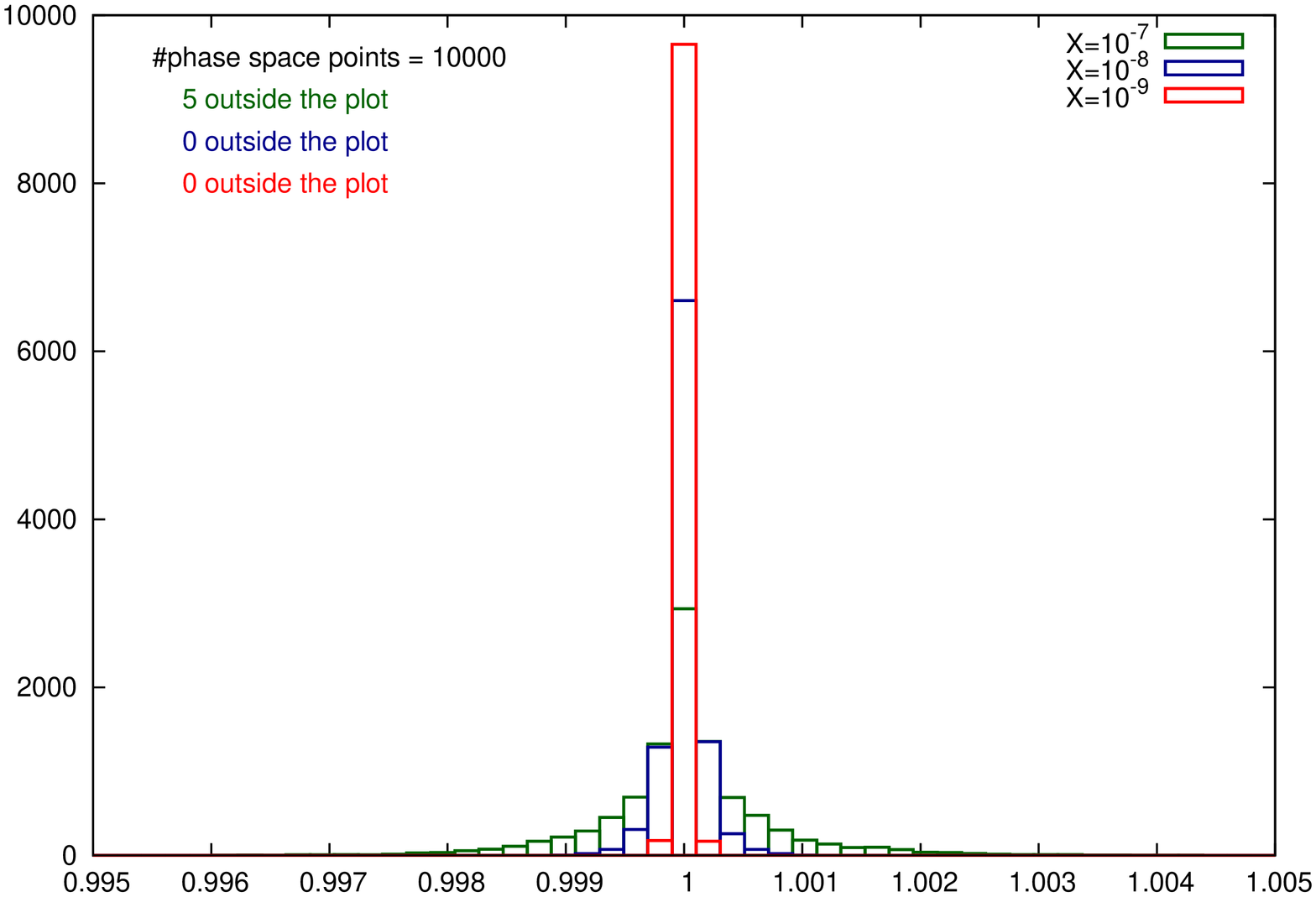}\\
\vspace{0.2cm}
(a)
\end{minipage}
\hspace{0.5cm}
\begin{minipage}[b]{0.7\linewidth}
\centering
\includegraphics[width=5cm,angle=270]{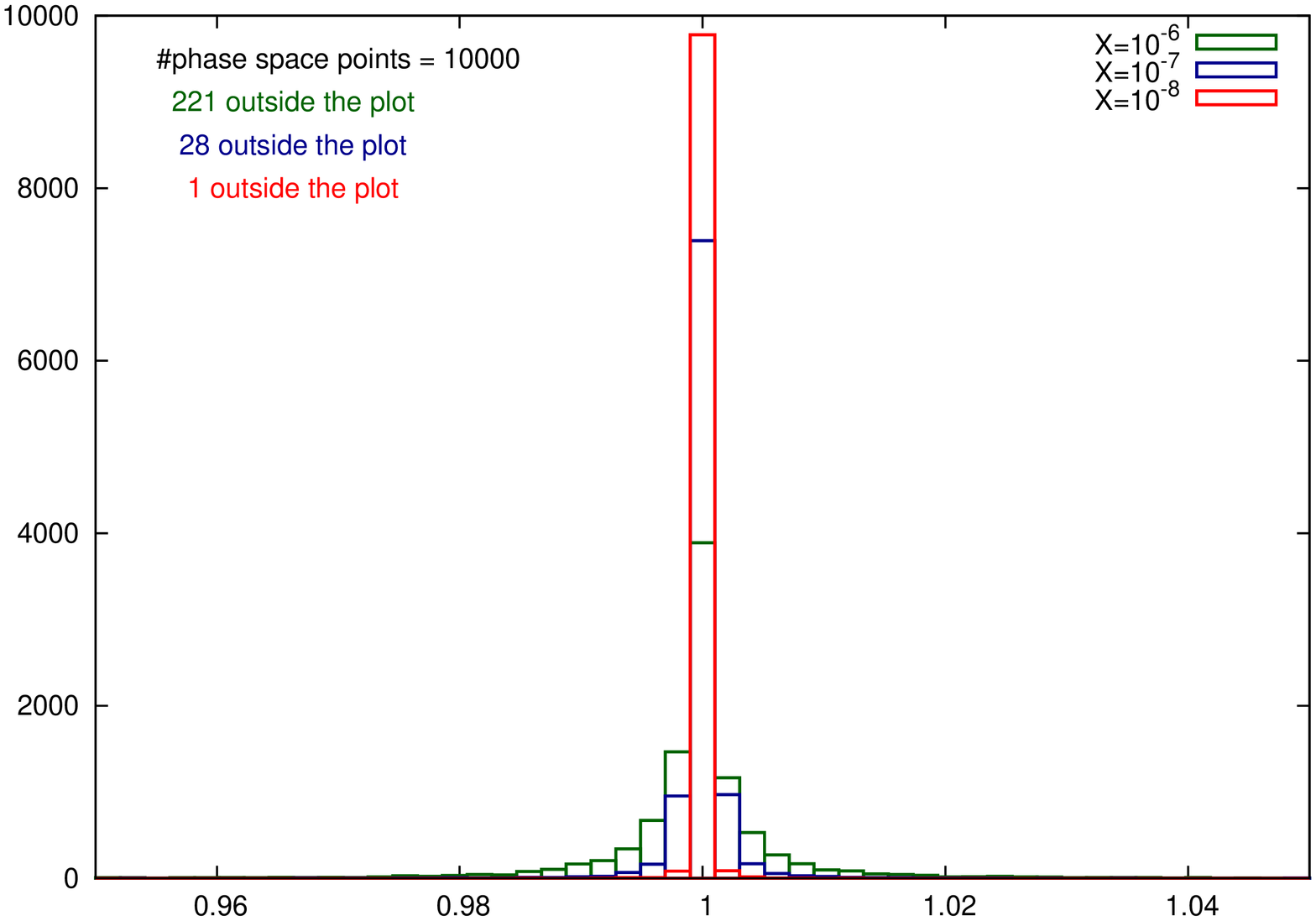}\\
\vspace{0.2cm}
(b)
\end{minipage}
\caption[Single collinear limit final state singularity]{Distribution of $R_{RV}$ for 10000 single collinear phase space point pairs where the pair of phase space points is related by an azimuthal rotation of $\pi/2$ about the collinear direction. (a) Final state collinear singularity 
(b) Initial state collinear singularity}
\label{fig:col-rot}
\end{figure}

We also tested how well the real-virtual subtraction term $\dsigma_{NNLO}^T$ approaches the real-virtual contribution $\dsigma_{NNLO}^{RV}$ in all single unresolved
regions of the phase space so that their difference can be integrated numerically in four dimensions. This is shown in fig.~\ref{fig:col-rot} for the single collinear case.
We noted~\cite{GehrmannDeRidder:2011aa} that in the real-virtual channel the magnitude of the angular correlations is much smaller than in the double-real case. Nevertheless
the combination of two phase space points with azimuthal angles  $\phi$ and $\phi+\frac{\pi}{2}$ can be used to provide an improvement in the convergence of the 
subtraction term.

\subsection{Double-virtual contribution}
For the double-virtual contribution we are concerned with the NNLO contribution coming from two-loop
processes,  $\dsigma_{NNLO}^{VV}$, the remaining integrated subtraction terms and NNLO mass factorisation contributions that are collectively denoted $\dsigma_{NNLO}^{U}$ in (\ref{eq:Udef}).
We note that this piece includes the full integrated real-virtual subtraction term $\dsigma_{NNLO}^{VS}$ derived in~\cite{GehrmannDeRidder:2011aa}  as well as 
integrated contributions from the double-real subtraction term
derived in~\cite{Glover:2010im}. This involves the integrated initial-initial four parton antenna functions which were recently computed in~\cite{iiNNLOantennae}, the integrated one-loop
three parton antennae~\cite{Gehrmann:2011wi} and 
convolutions of integrated three parton antennae~\cite{VV}.
We have verified that explicit poles from the double-virtual matrix elements cancel analytically against poles from this subtraction term. More explicitly~\cite{VV},
\begin{equation}
\Poles\left({\rm d}\hat{\sigma}_{NNLO}^{VV}+\int_2{\rm d}\hat{\sigma}_{NNLO}^{S,2}+\int_1{\rm d}\hat{\sigma}_{NNLO}^{VS}+{\rm d}\hat{\sigma}_{NNLO}^{MF,2}\right)=0\nonumber
\end{equation}
where the operator $\Poles$ selects the singular contributions containing $\epsilon$-poles that start at ${\cal O}(\epsilon^{-4})$. This gives us strong confidence on the derivation
of the subtraction terms in~\cite{Glover:2010im,GehrmannDeRidder:2011aa} and on the correctness of the results presented in~\cite{iiNNLOantennae}. 
In addition, it tells us that the infrared singular structure of the pure gluon scattering matrix elements at NNLO is captured analytically by the antenna subtraction method in a systematic and accurate manner.

\section{Conclusions}
In this talk we discussed the generalisation of the antenna subtraction method for the calculation of NNLO QCD corrections for exclusive collider observables with coloured partons
in the initial state. We considered the pure gluon channel contributing to the dijet cross section at NNLO and showed that the explicit $\epsilon$-poles in real-virtual and double-virtual
contributions are analytically cancelled by the $\epsilon$-poles in the subtraction terms. The finite remainders in the double-real, real-virtual and virtual-virtual channels can be
evaluated numerically in four dimensions by generating appropriate weighted partonic events.

Future steps include the construction of a numerical program to compute NNLO QCD corrections to dijet production in hadron-hadron collisions with the inclusion of the remaining
channels involving quark initiated processes \cite{Currie:2011nb}.

\section{Acknowledgements}
The research work reported in this talk has been done in collaboration with Aude Gehrmann-De Ridder, Thomas Gehrmann and Nigel Glover.
This research was supported by the Swiss National Science Foundation (SNF) under contract PP00P2-139192 and in part by the 
European Commission through the 'LHCPhenoNet' Initial Training Network PITN-GA-2010-264564', which are hereby acknowledged.

\end{document}